%% file: ufo_HESS.tex
\documentclass[a4paper,11pt]{article}
\usepackage{pos}
\usepackage{graphicx}
\usepackage{epsfig} 
\usepackage{bm} 
\usepackage{amsmath}
\usepackage{amssymb} 
\usepackage{color} 
\usepackage{float}
\usepackage{hyperref}
\usepackage{wasysym}
\usepackage{wrapfig, xspace, lipsum, todonotes}

\def\flat{\textit{Fermi}-LAT\xspace}
\def\hess{H.E.S.S.\xspace}

\title{Search for dark matter annihilation signals from UFOs with H.E.S.S.}

 \ShortTitle{ UFO's search with H.E.S.S.}

\author*[a]{D. Malyshev}
\author*[b]{A. Montanari}
\author*[b]{E. Moulin}
\author[c]{D. Glawion}

\forColl{H.E.S.S.}
\affiliation[a]{Institut f\"ur Astronomie und Astrophysik, Universit\"at T\"ubingen, Sand 1, D 72076 T\"ubingen, Germany}
\affiliation[b]{IRFU, CEA, Universit\'e Paris-Saclay,
  F-91191 Gif-sur-Yvette, France}
\affiliation[c]{ ECAP, FAU Erlangen-Nürnberg, D-91058, Erlangen, Germany}

\emailAdd{denys.malyshev@astro.uni-tuebingen.de}
\emailAdd{alessandro.montanari@cea.fr}
\emailAdd{emmanuel.moulin@cea.fr}
\emailAdd{dorit.glawion@fau.de}

\abstract{
Cosmological N-body simulations show that Milky-Way-sized galaxies harbor a population of unmerged dark matter subhalos. These subhalos could shine in gamma rays and be eventually detected in gamma-ray surveys as unidentified sources. We search for very-high-energy (VHE, $E\geq 100$~GeV) gamma-ray emission using H.E.S.S. observations carried out from a thorough selection of unidentified Fermi-LAT Objects (UFOs) as dark matter subhalo candidates. Provided that the dark matter mass is higher than a few hundred GeV, the emission of the UFOs can be well described by dark matter annihilation models. No significant VHE gamma-ray emission is detected in any UFO dataset nor in their combination. We, therefore, derive constraints on the product of the velocity-weighted annihilation cross-section $\langle \sigma v\rangle$ by the $J$-factor on dark matter models describing the UFO emissions. Upper limits at 95\% confidence level are derived on $\langle \sigma v\rangle J$ in $W^+W^-$ and $\tau^+\tau^-$ annihilation channels for the TeV dark matter particles. Focusing on thermal WIMPs, strong constraints on the $J$-factors are obtained from H.E.S.S. observations. Adopting model-dependent predictions from cosmological N-body simulations on the $J$-factor distribution function for Milky Way (MW)-sized galaxies, only $\lesssim 0.3$~TeV mass dark matter models marginally allow to explain observed UFO emission.}

\FullConference{37$^{\rm{th}}$ International Cosmic Ray Conference (ICRC 2021)\\
		July 12th -- 23rd, 2021\\
		Online -- Berlin, Germany}

\begin{document}
\maketitle

\section{Introduction}
\label{sec:intro}
Although the presence of dark matter (DM) in a variety of astrophysical objects is supported by a wealth of 
observations, its underlying microscopic nature is still unknown. One of the most popular particle-physics DM candidates is a weakly interacting massive particle (WIMP). Thermally-produced in the early universe with mass and coupling strength at the electroweak scale, these particles can make a present-day DM density~\citep{Steigman:2012nb} consistent with observations~\citep{Adam:2015rua}. Gamma-rays produced in the WIMP self-annihilation process for a long time have been recognized as prime messenger for indirect DM searches. The most compelling constraints on the parameters of annihilating TeV mass-scale WIMPs are derived from non-detection of the signal by ground-based imaging atmospheric Cherenkov telescopes (IACTs) such as \hess from the Galactic Centre~\citep{Abdallah:2016ygi,Abdallah:2018qtu} and nearby dwarf
galaxies~\citep{Aharonian:2007km,Abramowski:2010aa,Abramowski:2014tra,Abdalla:2018mve,Abdallah:2020sas}.

Another promising and complementary targets for indirect DM searches are DM subhalos populating the Galactic halo~\citep[see, e.g.,][]{Kamionkowski:2010mi}. The smallest structures are believed to have formed first in the observed Universe. Gravitationally-bound systems 
are formed by the collapse of DM particles. These systems later merge to form the first subhalos, which subsequently form more massive ones. The merging history leads to DM halos massive enough to retain gas and trigger star formation and give rise to the formation of galaxies we observe today. At the same time, most of the subhalos may not host significant amount of baryonic matter which makes them invisible at all wavelengths. However, in case of self-annihilating WIMP nature of the DM, subhalos could shine in gamma rays. The annihilation process of massive enough WIMPs in subhalos could be frequent enough to be detectable at GeV/TeV energies.
Given the unknown actual location of most of the DM subhalos, their searches can be performed using all-sky gamma-ray observations~\citep[see, e.g.,][]{Diemand:2006ik} such as with the Large Area Telescope (LAT) instrument onboard the Fermi satellite~\citep[see, for instance,][]{Berlin:2013dva} 
or wide-field surveys carried out with IACTs~\citep[see, for instance,][]{Aharonian:2008wt,Brun:2010ci}.


All-sky \flat observations revealed a significant population of sources lacking firm associations at other wavelengths~\citep{TheFermi-LAT:2017pvy,Fermi-LAT:2019yla}. These sources are therefore classified as unidentified Fermi objects (UFOs). The possible annihilating-WIMP dark matter origin of some of these objects was studied in~\citep{Belikov:2011pu,Zechlin:2011kk,Bertoni:2015mla,Bertoni:2016hoh,Calore:2016ogv,Coronado-Blazquez:2019puc} assuming relatively light WIMPs with masses below 100 GeV. At the same time the sub-population of UFOs characterized by a relatively hard spectrum without cutoff signatures in the GeV band can be good candidates for DM halos made of more massive ($\gtrsim 100$~GeV) WIMPs.
Such objects are therefore excellent targets for IACTs to perform searches for TeV DM subhalos. In 2018 and 2019, the H.E.S.S. collaboration carried out an observational campaign for a selection of the most promising UFOs in order to probe their potential TeV-mass DM-induced emission. 


\section{Targets selection and data analysis}
\label{sec:data}
\subsection{Targets selection}
\label{sec:selection}
The best DM subhalo candidates for \hess observations among the unidentified \flat sources are determined through a thorough selection in the Third Catalog of High-Energy \flat Sources (3FHL)~\citep{TheFermi-LAT:2017pvy},
which includes pointlike sources detected above 10 GeV. The source selection requires: \textit{(i)} the unidentified sources to be steady (according to 3FHL catalogue\footnote{While the criterium on the variability provides steady candidates as expected for DM sources, \flat photon properties at the highest energies have been checked. None of them could be attributed to flaring of nearby \flat sources.}); \textit{(ii)}: exhibit a hard power-law spectral index ($\Gamma < 2$), as expected for DM-induced signals for DM masses above 100 GeV; \textit{(iii)} have no obvious conventional counterpart at other wavelengths; \textit{(iv)} be located at $>5^\circ$ off the galactic plane (to avoid potential contamination from foreground Galactic diffuse emission). 
The multi-wavelength (MWL) search for possible counterparts is based on the Fermi-LAT source coordinates in catalogs of MWL facilities (\textit{XMM-Newton}, \textit{ROSAT}, \textit{SUZAKU}, \textit{CGRO}, \textit{Chandra}, \textit{Swift}, \textit{WMAP}, \textit{RXTE}, \textit{Nustar}, \textit{SDSS}, \textit{Planck}, \textit{WISE}, \textit{HST}).
In addition, we require that selected sources are located at preferable \hess sky regions, \textit{i.e.} can be observed by \hess with a maximum zenith angle of $45^\circ$.

The selection criteria were applied on the 3FHL source catalog and resulted in a selection of only three UFOs. The basic information on the three UFOs is summarized in Tab.~\ref{tab:table2}.
\begin{table}[htb!]
\centering
{\scriptsize
\begin{tabular}{l | c | c | c | c | c | c| c |c}
\hline
\hline
Name & RA & Dec. & TS for & Position  & Pivot & Flux & Power-law  & $E_{\rm cut}$\\
    &   &  & $E\geq10$~GeV  & uncertainty & energy & at pivot energy  &  index & (95\% c.l.) \\
          & [degrees] & [degrees] &  &[arcmin] & [GeV]  & [$10^{-13}$ TeV\,cm$^{-2}$s$^{-1}$]  & &[GeV]\\
\hline
3FHL J0929.2-4110 & 142.3345 &  -41.1833 & 36 & 2.4 & 0.39   & $0.12\pm 0.01$ & $1.37\pm 0.07$ &  $>33$\\
3FHL J1915.2-1323$^\dagger$ & 288.8182 &  -13.3916 &23 & 3.0 & 62.8  &$2.1\pm 0.9$ & $1.5\pm 0.4$ &  $>35$\\
3FHL J2030.2-5037 & 307.5901 &  -50.6344 & 40 & 2.6 & 6.3 & $1.9\pm 0.3$& $1.85\pm 0.1$& $>67$\\
\hline
\hline
\end{tabular}
}
\caption{\label{tab:table2} Properties of the selected UFOs together with their spectral parameters in $>0.1$~GeV band. The columns summarize RA-Dec coordinates of the UFOs, their test statistics values(TS), position uncertainty, pivot energy, best-fit flux at the pivot energy, power-law spectral index and 95\% c.l. lower limit on the cut-off energy. 
The 3FHL J1915.2-1323 source marked with $^\dagger$ is detected only above 10~GeV. For this source all spectral parameters are given for this energy band. }
\end{table}
\subsection{Expected signal}
\label{sec:signal}
The differential gamma-ray flux from Majorana DM particles of mass $m_{\rm DM}$ self-annihilating in object of size $\Delta \Omega$ is
\begin{equation}
\label{eq:dmflux}
\frac{{\rm d} \Phi_\gamma}{{\rm d} E_\gamma}  (E_\gamma, \Delta\Omega)=
\frac {\langle \sigma v \rangle}{8\pi m_{\rm DM}^2}\sum_f  \text{BR}_f \frac{{\rm d} N^f}{{\rm d}E_\gamma} \,J(\Delta\Omega) \ ,\,\,
{\rm with}\, J(\Delta\Omega) =  \int\limits_{\Delta\Omega}\int\limits_{\rm l.o.s.}
\rho^2(s(r,\theta)) dsd\Omega\,  .
\end{equation}
$\langle \sigma v \rangle$ is the thermally-averaged  velocity-weighted annihilation cross section and $\sum\limits_f  \text{BR}_f  dN^f/dE_{\gamma}$ is the sum of the annihilation spectra ${\rm d}N^f/{\rm d}E_{\gamma}$ per annihilation in the final states $f$ with associated branching ratios ${\rm BR}_{f}$.
Hereafter we will refer to the quantity $J(\Delta\Omega)$ as a total $J$-factor within a solid angle $\Delta\Omega$. We note, that for point-like DM subhalos the strength of the DM annihilation signal is proportional to the quantity $\langle \sigma v \rangle J$, where $J$ -- is a $J$-factor of clump integrated over the point spread function of the instrument. In this case detection or non-detection of the annihilation signal allows one to directly measure or constrain this quantity.

As opposed to objects with measured stellar dynamics like dwarf galaxies, UFOs have unknown distances to Earth and their $J$-factors cannot be derived from stellar kinematics. To access the proper values of $J$-factors for the selected DM subhalos in what follows, we adapt a statistical approach based on utilising subhalos' $J$-factor distribution as seen in $N$-body cosmological simulations (see, for instance, Refs.~\citep{Diemand:2008in,Springel:2008cc}). In order to derive the $J$-factor distribution of DM subhalos in the MW, we used the CLUMPY code v3.0.0~\citep{clumpy,clumpy_v2,clumpy_v3} and performed  
1000 simulations of a MW-like galaxy with a smooth NFW~\citep{Navarro:1996gj} DM main halo profile with the parameters corresponding to the best-fit NFW parameters from Ref.~\citep{cautun20}. 
For each simulation, the subhalo parameters were chosen similar to the ones used in~\citep{hutten16} for the ``HIGH'' model which results in somewhat optimistic values of the obtained $J$-factors.
From each simulation we derived the Galactic coordinates of all subhalos and their $J$-factors integrated in circular regions with 0.1$^\circ$.

\begin{figure*}
\begin{center}
\includegraphics[width=0.5\linewidth]{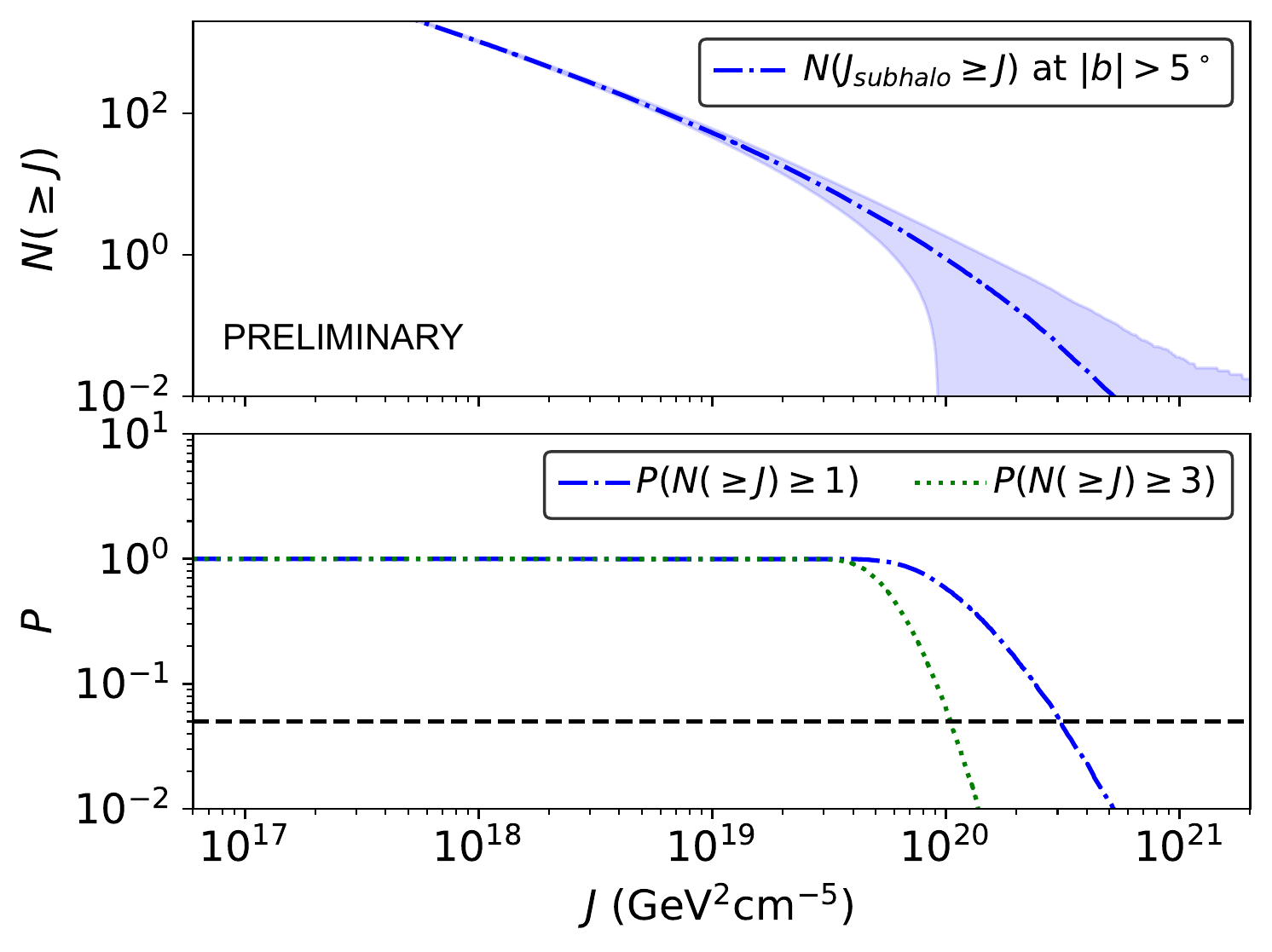}
\caption{{\it Top panel:} Cumulative $J$-factor distribution, $N(\ge J)$, of a MW-like subhalo population. The number of subhalos with a $J$-factor exceeding a given value is plotted (blue dot-dashed curve).
The blue-shaded band corresponds to the 1$\sigma$ statistical uncertainty. 
{\it Bottom panel:}  
Probability $P$ to find at least one (three) subhalo(s) with a $J$-factor higher than specified are shown with blue(green) lines. The horizontal black dashed line shows the 5\%  probability. The figure was adapted from~\cite{we_ufos}. See text for more details.}
\label{fig:luminosty_function}
\end{center}
\end{figure*}

The cumulative $J$-factor distribution $N(\geq J)$ is shown in the upper panel of Fig.~\ref{fig:luminosty_function} for subhalos located at Galactic latitudes $|b| > 5^\circ$ similar to the UFOs considered for the analysis. 
The dot-dashed blue curve shows the averaged distribution computed from all the realizations and the shaded region shows the formal 1$\sigma$ statistical dispersion calculated over all simulated MW-like galaxies. In the lower panel of Fig.~\ref{fig:luminosty_function}, the blue dot-dashed/green-dotted curves illustrate the probability to find in any simulation at least one/three subhalos with a $J$-factor higher than specified. The horizontal black-dashed line illustrates the 5\% probability. 
We conclude, that the the $J$-factor of one DM subhalo at 95\% c.l. can be constrained as $J\leq 3\cdot 10^{20}$~GeV$^2$cm$^{-5}$. The average $J$-factor of three subhalos is $J \leq 1\cdot 10^{20}$~GeV$^2$cm$^{-5}$.

\subsection{Data Analysis}
\label{sec:data_analysis}
\subsection{\flat data analysis }
\label{sec:fermi_data_analysis}
\flat data selected for the analysis spans for more than 12 years (Aug. 2008 to Oct. 2020). The data were analysed with \texttt{fermitools} v.~2.0.0  with P8R3\_V3 response functions (\texttt{CLEAN} photon class). We performed standard binned analysis of the data in $14^\circ$-radius region around positions of each UFO in 0.1-1000~GeV energy bin including to the model all sources from 4FGL-DR2 ctalogue~\citep{Fermi-LAT:2019yla} and standard diffuse background templates. The results of the modelling of UFOs with a cutoff powerlaw spectral model are summarized in Tab.~\ref{tab:table2}. For the rest of the presented results we explicitly assumed, that UFOs' spectra follow Eq.~\ref{eq:dmflux}.

\subsubsection{ \hess data analysis}
H.E.S.S. is an array of five IACTs located in the Khomas Highland in Namibia, at an altitude of 1800 m. The array is composed of four 12~m diameter telescopes (CT1-4)  and a fifth 28~m diameter telescope (CT5) at the middle of the array. 
The observations presented here were performed in 2018 and 2019 in the \textit{wobble} mode with the full five-telescope array for the selection of UFOs presented in Tab.~\ref{tab:table2}. The standard run selection criteria are applied to select the observations for the data analysis~\citep{Aharonian:2006pe}. The gamma-ray events direction and energy reconstruction is performed with a template-fitting technique~\citep{2009APh32231D}, after the calibration of raw shower images recorded in the camera. 

The selected UFOs are assumed to be pointlike sources according to the point spread function (PSF) of \flat which reaches $\sim$0.1$^\circ$ above 0.1~TeV. The region of interest (ROI), hereafter referred to as the ON source region, is defined, given the \hess PSF, as for pointlike-emission searches for H.E.S.S.. The ROI is taken as a disk of $0.12^\circ$ radius. For the measurement of the residual background in OFF regions, the {\it MultipleOff} technique~\citep{Aharonian:2006pe} is used. The excess significance in the ROI is computed following the statistical approach of Ref.~\citep{1983ApJ...272..317L}.
\begin{figure*}
\centering
\includegraphics[width=0.48\textwidth]{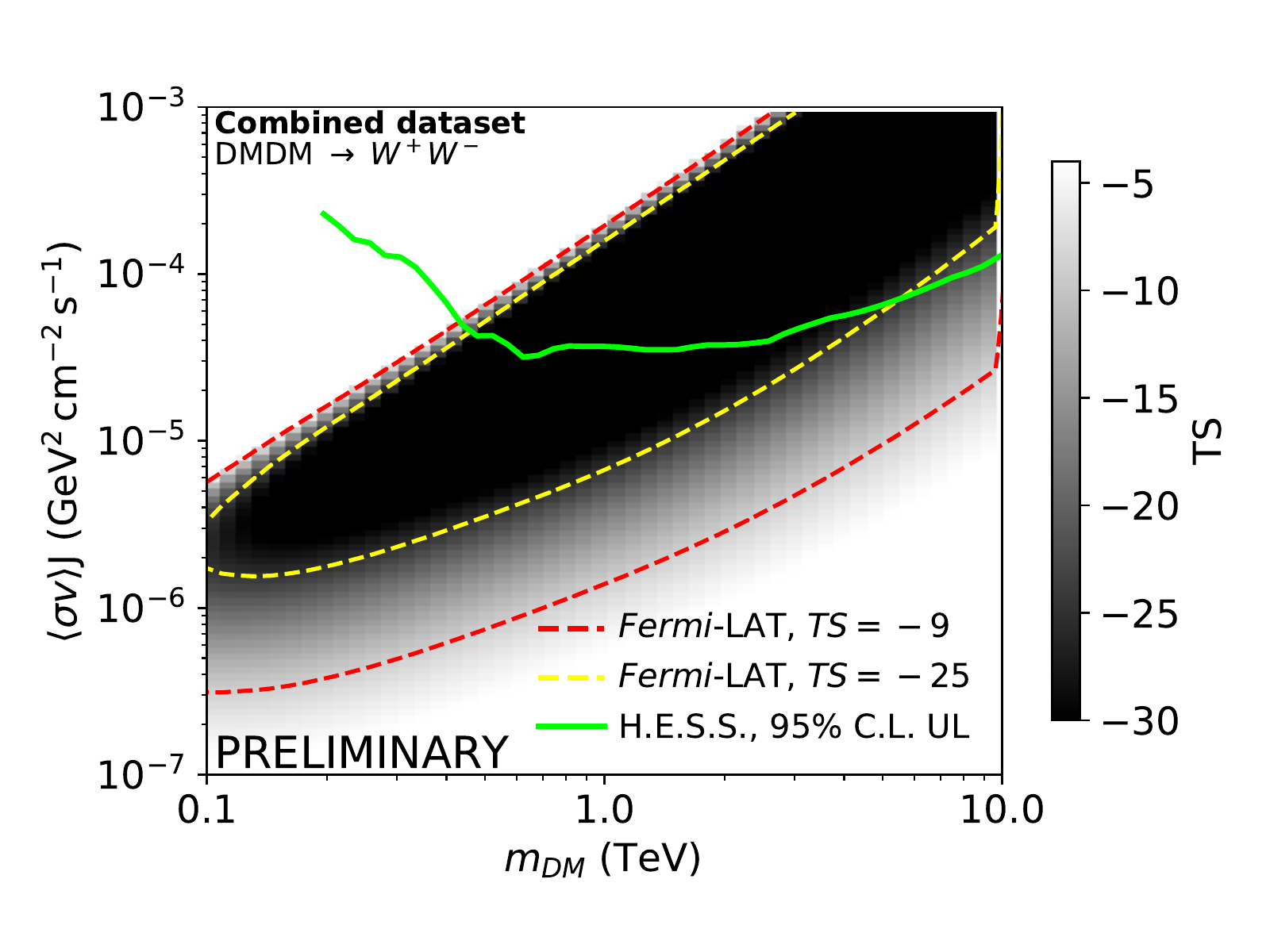}
\includegraphics[width=0.48\textwidth]{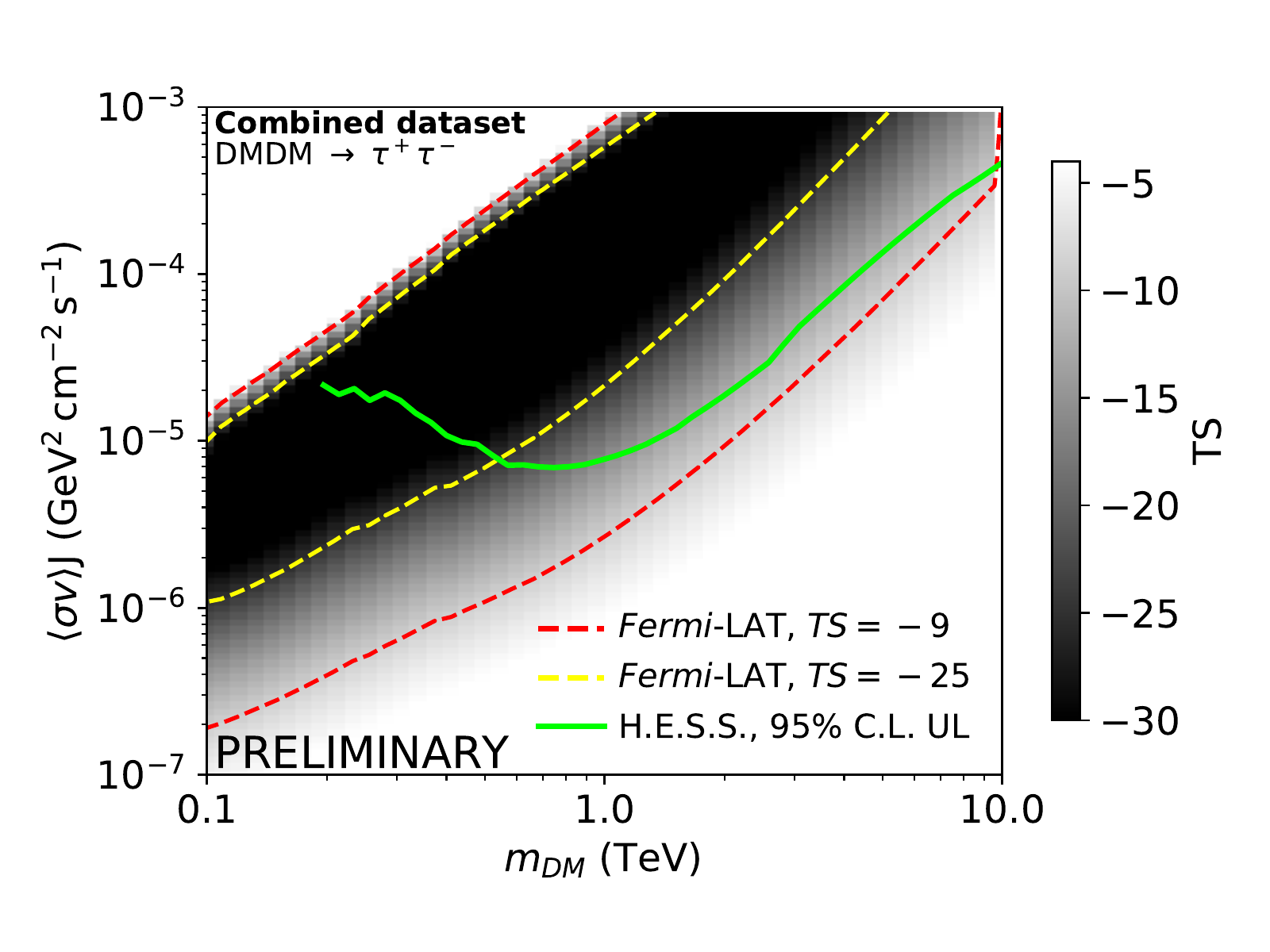}
\caption{Contours of $TS$ computed from \flat combined UFO datasets for $W^+W^-$ and $\tau^+\tau^-$ annihilation channels. The contours are given in the ($\langle \sigma v \rangle J$,m$_{\rm DM}$) plane. The cyan and orange dashed lines show the $-9$ and $-25$ $TS$ contours. Overlaid (solid green line) are H.E.S.S. upper limits displayed at 95\% C.L. The figure was adapted from~\cite{we_ufos}.}
\label{fig:FermiTSmap}
\end{figure*}

\section{Results}
\label{sec:results}
In absence of significant excess in any of the \hess datasets of the selected UFOs we provide 95\% c.l. upper limits (assuming the best-fit power-law spectral index in the \flat band) on the $\langle \sigma v \rangle J$ as a function of DM mass using 
using a log-likelihood ratio test statistic 
for the combined dataset of all UFOs, see Ref.~\citep{we_ufos} for the more details.
Figure~\ref{fig:FermiTSmap} shows derived upper limits as a function of the DM mass for the $W^+W^-$ and $\tau^+\tau^-$ annihilation channels, respectively. Green line presents \hess 95\% c.l. limits from the combined analysis of all UFOs, while the color presents the TS of the signal in \flat band. The combined limits reach 3.7$\times$10$^{-5}$ and 8.1$\times$10$^{-6}$ GeV$^2$cm$^{-2}$s$^{-1}$ in the $W^+W^-$ and $\tau^+\tau^-$ channels, respectively, for 1~TeV DM mass.

\begin{figure*}[htbp]
\centering
\includegraphics[width=0.48\textwidth]{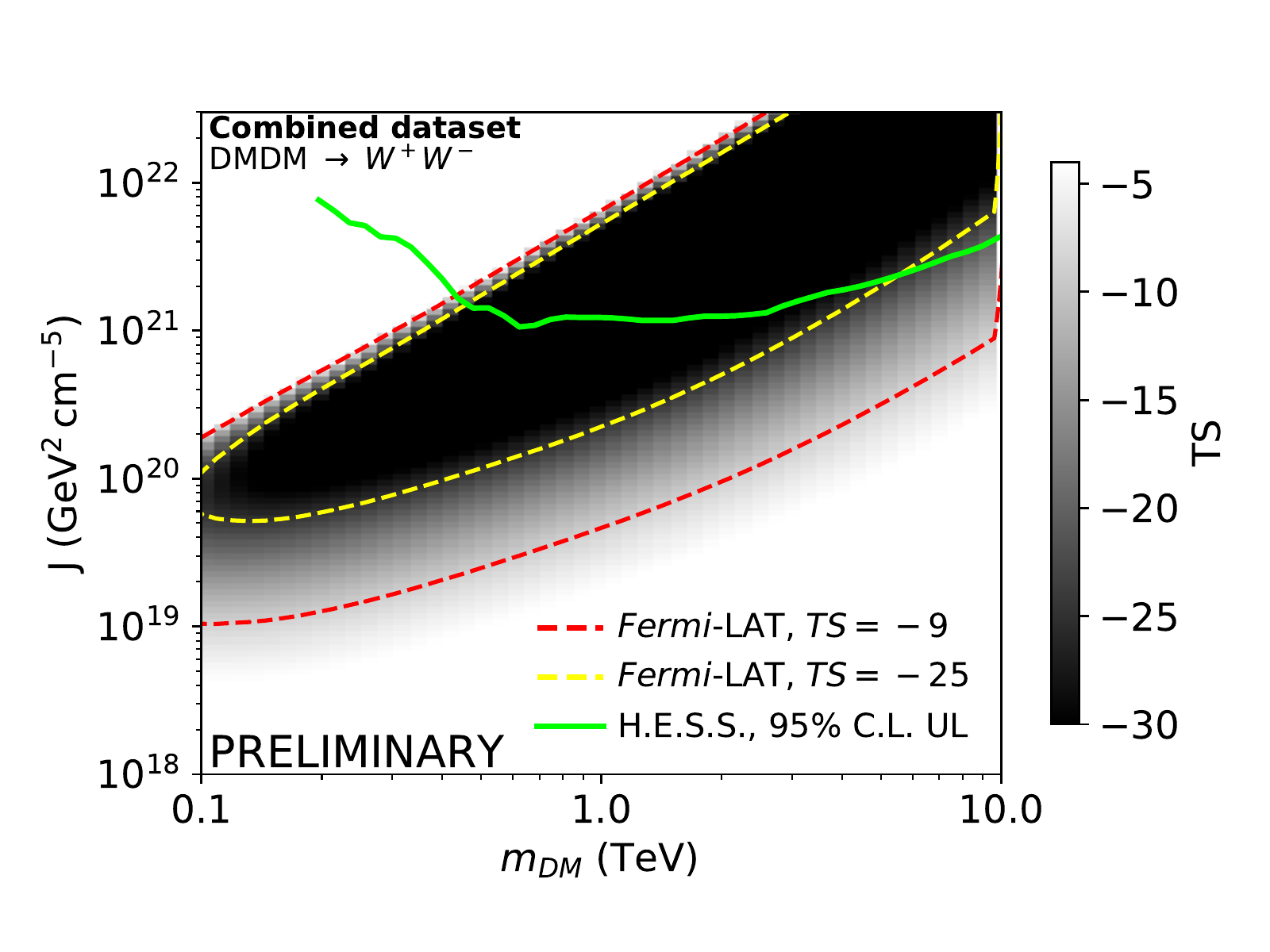}
\includegraphics[width=0.48\textwidth]{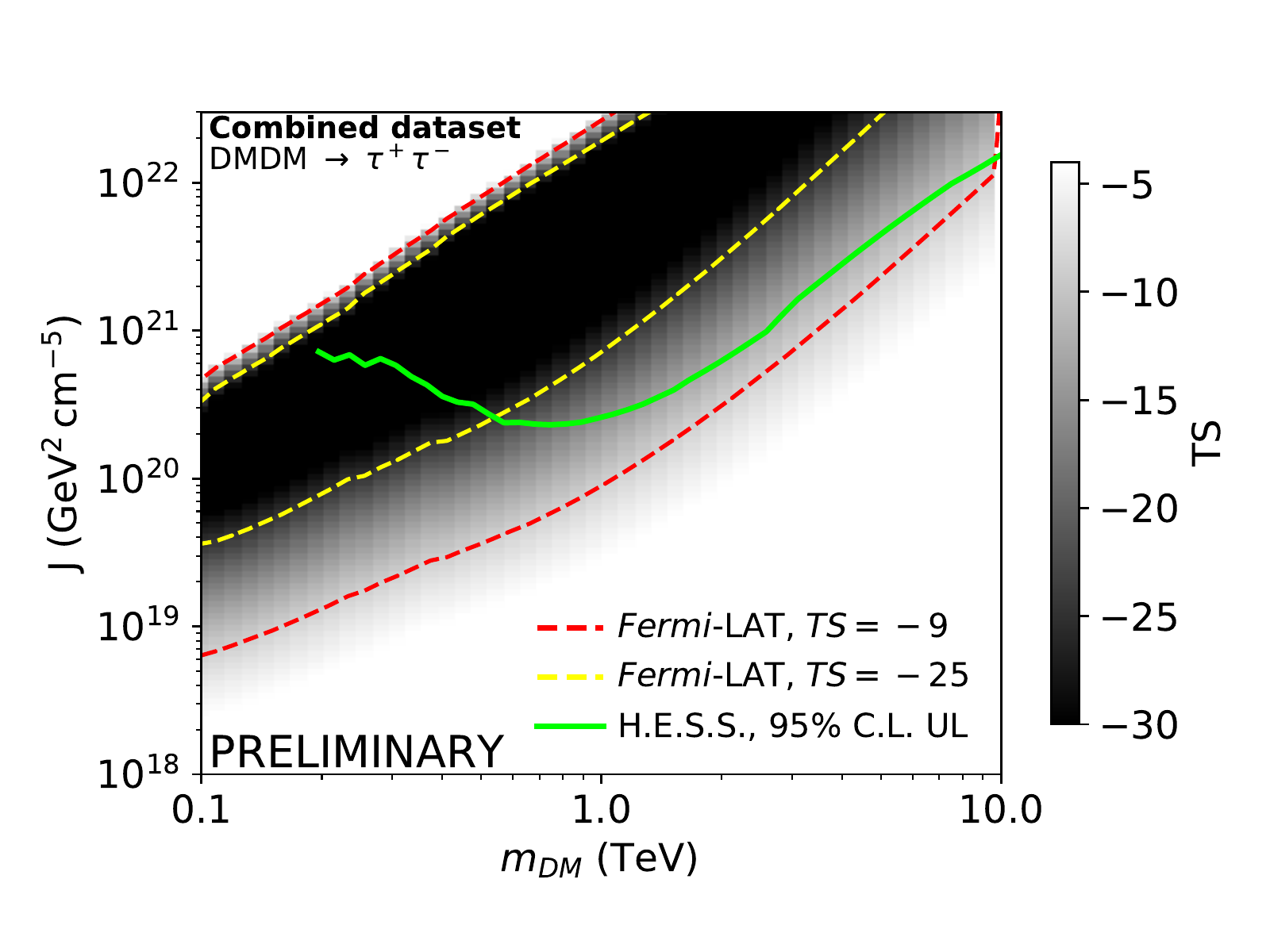}
\caption{ Contours of $TS$ computed from the \flat combined UFO datasets. The contours are given in the ($J$,m$_{\rm DM}$) plane for the $W^+W^-$ (left panel) and $\tau^+\tau^-$ (right panel) annihilation channel,
assuming the $\langle\sigma v\rangle$ value expected for thermal WIMPs.
The cyan and orange dashed lines show the $-$9 and $-$25 $TS$ contours. Overlaid (solid green line) are the H.E.S.S. 95\% C.L. upper limits from the combined UFO datasets. The figure was adapted from~\cite{we_ufos}. }
\label{fig:sigmavJ_stacked}
\end{figure*}

In order to derive the $J$-factor values required to explain the UFO emission in terms of DM models, the value of the annihilation cross section expected for thermal WIMPs ($\langle\sigma v\rangle_{\rm th} \simeq$ 3$\times$10$^{-26}$ cm$^3$s$^{-1}$) is used~\citep{Steigman:2012nb}.
The $J$-factor upper limits for the DM models of the UFOs as function of the DM mass are given at 95\% C.L. in  Fig.~\ref{fig:sigmavJ_stacked}.
For a 1~TeV DM mass in the $W^+W^-$ channel, the $J$-factor values are constrained to be between $(0.24-1.3)\times$10$^{21}$ GeV$^2$cm$^{-5}$ for DM models with $TS \le$ $-$25 (which corresponds to $\ge$5$\sigma$ confidence interval assuming $TS$ follows $\chi^2$ distribution). For a DM mass of 10 TeV in the $W^+W^-$ channel, all the $J$-factor values for DM models with TS $\le$ $-$25 are ruled out at 95\% C.L. by the H.E.S.S. constraints. In the $\tau^+\tau^-$ channels, the H.E.S.S. constraints are even stronger. For 300~GeV DM mass, the allowed $J$-factor values are between 1.4$\times$10$^{20}$ and 5.9$\times$10$^{20}$ GeV$^2$cm$^{-5}$ for TS $\le$ $-$25 DM models. The H.E.S.S. upper limits restrict the J-factors to lie in the range $6.1\times 10^{19} -2.0\times 10^{21}$~GeV$^2$cm$^{-5}$ and the masses 
to lie between 0.2 and 6 TeV in the W$^+$W$^-$channel. For the $\tau^+\tau^-$ channel, the J-factors lie in the range $7.0\times 10^{19} - 7.1\times 10^{20}$~GeV$^2$cm$^{-5}$ and the masses lie between 0.2 and 0.5~TeV.

Using predictions of N-body cosmological simulations, the number of subhalos with a $J$-factor higher than a given value for a MW-like galaxy can be extracted as displayed in Fig.~\ref{fig:luminosty_function}. The probability to have at least three subhalos with a $J$-factor higher than 10$^{20}$ GeV$^2$cm$^{-5}$ is below 5\%. 
According to this prediction, the interpretation of the UFO emissions in terms of DM particle annihilations in Galactic DM subhalos can be further
constrained from Fig.~\ref{fig:sigmavJ_stacked} to $m_{\rm DM}\lesssim 1$~TeV for $W^+W^-$ and $m_{\rm DM}\lesssim 0.3$~TeV for $\tau^+\tau^-$ channels. 

\section{Discussion and conclusions}
An important number of UFOs may produce gamma-rays from the annihilation process of DM in subhalos. However, some of them could be AGNs or other types of galaxies still lacking a detection at other wavelengths. Less plausible alternative astrophysical interpretations of UFOs could be as pulsars or low-luminosity globular clusters hosting millisecond pulsars~\citep{Mirabal:2016huj}. However, the energy cut-off of the gamma-ray spectra for these types of objects consists of a few GeV. The cumulative $J$-factor distribution is in very good agreement with the results of Ref.~\citep{hutten16} for the "HIGH" model intended to predict the highest possible number of subhalos in a typical MW-like galaxy. As it is shown fro the predictions in the ``LOW'' model of Ref.~\citep{hutten16}, the real number of DM subhalos can be an order of magnitude smaller.
The output of DM-only simulations dictates the choice of the number of subhalos of masses between 10$^8$ and 10$^{10}$ M$_\odot$ of $N_{\rm calib}$ = 300~\cite{Springel:2008cc}. The latter can be significantly reduced (up to a factor of two) by baryon feedback~\citep{Mollitor:2014ara,Sawala:2015cdf}. As a consequence, the highest J-factor values would be even more unlikely. As discussed in ref.~\citep{Coronado-Blazquez:2019pny}, \flat should observe the subhalos with the highest $J$-factors as extended scources, given the point spread function of about 0.1$^\circ$ above 10~GeV for the Fermi satellite. However, even the gamma-ray sources produced by these brightest DM subhalos would be faint. The spatial extension of these sources would be challenging to measure for \flat. On the simulation front, 
further work is likely needed to use predictions for subhalo angular sizes in MW-like galaxies to definitely rule out pointlike
UFOs as potential DM subhalos.

Interpreting UFOs as DM subhalos of TeV-mass scale thermal WIMPs requires $J$-factors to be larger than a few 10$^{20}$~GeV$^2$cm$^{-5}$. From the N-body simulations of MW-type galaxies, only occasionaly such $J$-factor values can be obtained. A large statistical variance affects the highest subhalo $J$-factor. The precise value of the brightest subhalos can be subject to a large uncertainty. A factor of ten uncertainty is implied for the $J$-factor value for J $\gtrsim$ 10$^{20}$ GeV$^2$cm$^{-5}$  in the "HIGH" model~\citep{hutten19}.
Additional factor of several uncertainties can be connected to the uncertainties of the DM distribution in the MW~\citep{Read:2014qva} and/or presence of substructures in the galactic subhalos \citep{Hiroshima:2018kfv}.
The constraints from cosmological simulations are significantly weakened by the above-mentioned large systematic uncertainties in the prediction of the $J$-factor distribution. This makes the former comparable to or weaker than the H.E.S.S. constraints in, {\it e.g.}, the $\tau^+\tau^-$ channel. Therefore the model-independent H.E.S.S. constraints are the only relevant and robust ones when interpreting the UFOs as Galactic subhalos of annihilating DM.

\noindent\textbf{Acknowledgements.}
H.E.S.S. gratefully acknowledges financial support from the agencies and organizations listed at \href{https://www.mpi-hd.mpg.de/hfm/HESS/pages/publications/auxiliary/HESS-Acknowledgements-2021.html}{H.E.S.S.-Acknowledgements webpage}.

\input{journals.tex}

\bibliography{bibl.bib} 
\appendix
\clearpage
\section*{Full author list}
H.~Abdalla$^{1}$, 
F.~Aharonian$^{2,3,4}$, 
F.~Ait~Benkhali$^{3}$, 
E.O.~Ang\"uner$^{5}$, 
C.~Arcaro$^{6}$, 
C.~Armand$^{7}$, 
T.~Armstrong$^{8}$, 
H.~Ashkar$^{9}$, 
M.~Backes$^{1,6}$, 
V.~Baghmanyan$^{10}$
% ...

\medskip

\noindent
$^{1}$University of Namibia, Department of Physics, Private Bag 13301, Windhoek 10005, Namibia\\
$^{2}$Dublin Institute for Advanced Studies, 31 Fitzwilliam Place, Dublin 2, Ireland\\
$^{3}$Max-Planck-Institut f\"ur Kernphysik, P.O. Box 103980, D 69029 Heidelberg, Germany\\
$^{4}$High Energy Astrophysics Laboratory, RAU,  123 Hovsep Emin St  Yerevan 0051, Armenia\\
$^{5}$Aix Marseille Universit\'e, CNRS/IN2P3, CPPM, Marseille, France\\
$^{6}$Centre for Space Research, North-West University, Potchefstroom 2520, South Africa\\
$^{7}$Laboratoire d'Annecy de Physique des Particules, Univ. Grenoble Alpes, Univ. Savoie Mont Blanc, CNRS, LAPP, 74000 Annecy, France\\
$^{8}$University of Oxford, Department of Physics, Denys Wilkinson Building, Keble Road, Oxford OX1 3RH, UK\\
$^{9}$IRFU, CEA, Universit\'e Paris-Saclay, F-91191 Gif-sur-Yvette, France\\
$^{10}$Instytut Fizyki J\c{a}drowej PAN, ul. Radzikowskiego 152, 31-342 Krak{\'o}w, Poland\\
% ...

%
%
%

\end{document}

%% file: journals.tex
\def\aj{AJ}%
\def\actaa{Acta Astron.}%
\def\araa{ARA\&A}%
\def\apj{ApJ}%
\def\apjl{ApJ}%
\def\apjs{ApJS}%
\def\ao{Appl.~Opt.}%
\def\apss{Ap\&SS}%
\def\aap{A\&A}%
\def\aapr{A\&A~Rev.}%
\def\aaps{A\&AS}%
\def\azh{AZh}%
\def\baas{BAAS}%
\def\bac{Bull. astr. Inst. Czechosl.}%
\def\caa{Chinese Astron. Astrophys.}%
\def\cjaa{Chinese J. Astron. Astrophys.}%
\def\icarus{Icarus}%
\def\jcap{J. Cosmology Astropart. Phys.}%
\def\jrasc{JRASC}%
\def\mnras{MNRAS}%
\def\memras{MmRAS}%
\def\na{New A}%
\def\nar{New A Rev.}%
\def\pasa{PASA}%
\def\pra{Phys.~Rev.~A}%
\def\prb{Phys.~Rev.~B}%
\def\prc{Phys.~Rev.~C}%
\def\prd{Phys.~Rev.~D}%
\def\pre{Phys.~Rev.~E}%
\def\prl{Phys.~Rev.~Lett.}%
\def\pasp{PASP}%
\def\pasj{PASJ}%
\def\qjras{QJRAS}%
\def\rmxaa{Rev. Mexicana Astron. Astrofis.}%
\def\skytel{S\&T}%
\def\solphys{Sol.~Phys.}%
\def\sovast{Soviet~Ast.}%
\def\ssr{Space~Sci.~Rev.}%
\def\zap{ZAp}%
\def\nat{Nature}%
\def\iaucirc{IAU~Circ.}%
\def\aplett{Astrophys.~Lett.}%
\def\apspr{Astrophys.~Space~Phys.~Res.}%
\def\bain{Bull.~Astron.~Inst.~Netherlands}%
\def\fcp{Fund.~Cosmic~Phys.}%
\def\gca{Geochim.~Cosmochim.~Acta}%
\def\grl{Geophys.~Res.~Lett.}%
\def\jcp{J.~Chem.~Phys.}%
\def\jgr{J.~Geophys.~Res.}%
\def\jqsrt{J.~Quant.~Spec.~Radiat.~Transf.}%
\def\memsai{Mem.~Soc.~Astron.~Italiana}%
\def\nphysa{Nucl.~Phys.~A}%
\def\physrep{Phys.~Rep.}%
\def\physscr{Phys.~Scr}%
\def\planss{Planet.~Space~Sci.}%
\def\procspie{Proc.~SPIE}%
\let\astap=\aap
\let\apjlett=\apjl
\let\apjsupp=\apjs
\let\applopt=\ao

%% file: ufo_HESS.bbl
\begin{thebibliography}{41}
\expandafter\ifx\csname natexlab\endcsname\relax\def\natexlab#1{#1}\fi

\bibitem[{Abdalla {et~al.}(2018)}]{Abdalla:2018mve}
Abdalla, H. {et~al.} 2018, JCAP, 11, 037

\bibitem[{Abdallah {et~al.}(2016)}]{Abdallah:2016ygi}
Abdallah, H. {et~al.} 2016, Phys. Rev. Lett., 117, 111301

\bibitem[{Abdallah {et~al.}(2018)}]{Abdallah:2018qtu}
Abdallah, H. {et~al.} 2018, Phys.\ Rev.\ Lett., 120, 201101

\bibitem[{Abdallah {et~al.}(2020)}]{Abdallah:2020sas}
Abdallah, H. {et~al.} 2020, Phys. Rev. D, 102, 062001

\bibitem[{Abdollahi {et~al.}(2020)}]{Fermi-LAT:2019yla}
Abdollahi, S. {et~al.} 2020, Astrophys. J. Suppl., 247, 33

\bibitem[{Abramowski {et~al.}(2011)}]{Abramowski:2010aa}
Abramowski, A. {et~al.} 2011, Astropart.\ Phys., 34, 608

\bibitem[{Abramowski {et~al.}(2014)}]{Abramowski:2014tra}
Abramowski, A. {et~al.} 2014, Phys. Rev., D90, 112012

\bibitem[{Adam {et~al.}(2016)}]{Adam:2015rua}
Adam, R. {et~al.} 2016, Astron. Astrophys., 594, A1

\bibitem[{Aharonian {et~al.}(2006)}]{Aharonian:2006pe}
Aharonian, F. {et~al.} 2006, Astron. Astrophys., 457, 899

\bibitem[{Aharonian {et~al.}(2008{\natexlab{a}})}]{Aharonian:2007km}
Aharonian, F. {et~al.} 2008{\natexlab{a}}, Astropart. Phys., 29, 55, [Erratum:
  Astropart.Phys. 33, 274--275 (2010)]

\bibitem[{Aharonian {et~al.}(2008{\natexlab{b}})}]{Aharonian:2008wt}
Aharonian, F. {et~al.} 2008{\natexlab{b}}, Phys. Rev. D, 78, 072008

\bibitem[{Ajello {et~al.}(2017)}]{TheFermi-LAT:2017pvy}
Ajello, M. {et~al.} 2017, Astrophys.\ J.\ Suppl., 232, 18

\bibitem[{Belikov {et~al.}(2012)Belikov, Hooper, \& Buckley}]{Belikov:2011pu}
Belikov, A.~V., Hooper, D., \& Buckley, M.~R. 2012, Phys. Rev. D, 86, 043504

\bibitem[{Berlin \& Hooper(2014)}]{Berlin:2013dva}
Berlin, A. \& Hooper, D. 2014, Phys. Rev. D, 89, 016014

\bibitem[{Bertoni {et~al.}(2015)Bertoni, Hooper, \& Linden}]{Bertoni:2015mla}
Bertoni, B., Hooper, D., \& Linden, T. 2015, JCAP, 12, 035

\bibitem[{Bertoni {et~al.}(2016)Bertoni, Hooper, \& Linden}]{Bertoni:2016hoh}
Bertoni, B., Hooper, D., \& Linden, T. 2016, JCAP, 05, 049

\bibitem[{{Bonnivard} {et~al.}(2016){Bonnivard}, {H{\"u}tten}, {Nezri},
  {Charbonnier}, {Combet}, \& {Maurin}}]{clumpy_v2}
{Bonnivard}, V., {H{\"u}tten}, M., {Nezri}, E., {et~al.} 2016, Computer Physics
  Communications, 200, 336

\bibitem[{Brun {et~al.}(2011)Brun, Moulin, Diemand, \&
  Glicenstein}]{Brun:2010ci}
Brun, P., Moulin, E., Diemand, J., \& Glicenstein, J.-F. 2011, Phys. Rev. D,
  83, 015003

\bibitem[{Calore {et~al.}(2017)Calore, De~Romeri, Di~Mauro, Donato, \&
  Marinacci}]{Calore:2016ogv}
Calore, F., De~Romeri, V., Di~Mauro, M., Donato, F., \& Marinacci, F. 2017,
  Phys. Rev. D, 96, 063009

\bibitem[{{Cautun} {et~al.}(2020){Cautun}, {Ben{\'\i}tez-Llambay}, {Deason},
  {Frenk}, {Fattahi}, {G{\'o}mez}, {Grand}, {Oman}, {Navarro}, \&
  {Simpson}}]{cautun20}
{Cautun}, M., {Ben{\'\i}tez-Llambay}, A., {Deason}, A.~J., {et~al.} 2020,
  \mnras, 494, 4291

\bibitem[{{Charbonnier} {et~al.}(2012){Charbonnier}, {Combet}, \&
  {Maurin}}]{clumpy}
{Charbonnier}, A., {Combet}, C., \& {Maurin}, D. 2012, Computer Physics
  Communications, 183, 656

\bibitem[{{Coronado-Bl{\'a}zquez} {et~al.}(2019){Coronado-Bl{\'a}zquez},
  {S{\'a}nchez-Conde}, {Di Mauro}, {Aguirre-Santaella}, {Ciuc{\u{a}}},
  {Dom{\'\i}nguez}, {Kawata}, \& {Mirabal}}]{Coronado-Blazquez:2019pny}
{Coronado-Bl{\'a}zquez}, J., {S{\'a}nchez-Conde}, M.~A., {Di Mauro}, M.,
  {et~al.} 2019, \jcap, 2019, 045

\bibitem[{Coronado-Blazquez {et~al.}(2019)Coronado-Blazquez, Sanchez-Conde,
  Dominguez, Aguirre-Santaella, Di~Mauro, Mirabal, Nieto, \&
  Charles}]{Coronado-Blazquez:2019puc}
Coronado-Blazquez, J., Sanchez-Conde, M.~A., Dominguez, A., {et~al.} 2019,
  JCAP, 07, 020

\bibitem[{{de Naurois} \& {Rolland}(2009)}]{2009APh32231D}
{de Naurois}, M. \& {Rolland}, L. 2009, Astropart. Phys., 32, 231

\bibitem[{Diemand {et~al.}(2007)Diemand, Kuhlen, \& Madau}]{Diemand:2006ik}
Diemand, J., Kuhlen, M., \& Madau, P. 2007, Astrophys. J., 657, 262

\bibitem[{Diemand {et~al.}(2008)Diemand, Kuhlen, Madau, Zemp, Moore, Potter, \&
  Stadel}]{Diemand:2008in}
Diemand, J., Kuhlen, M., Madau, P., {et~al.} 2008, Nature, 454, 735

\bibitem[{{H.~E.~S.~S. Collaboration} {et~al.}(2021){H.~E.~S.~S.
  Collaboration}, {Abdallah}, {Aharonian}, {Ait Benkhali}, {Ang{\"u}ner},
  {Arcaro}, {Armand}, {Armstrong}, {Ashkar}, {Backes}, {Baghmanyan}, {Barbosa
  Martins}, {Barnacka}, {Barnard}, {Becherini}, {Berge}, {Bernl{\"o}hr}, {Bi},
  {B{\"o}ttcher}, {Boisson}, {Bolmont}, {de Bony de Lavergne}, {Breuhaus},
  {Brun}, {Brun}, {Bryan}, {B{\"u}chele}, {Bulik}, {Bylund}, {Caroff},
  {Carosi}, {Casanova}, {Chand}, {Chandra}, {Chen}, {Cotter}, {Curylo},
  {Damascene Mbarubucyeye}, {Davids}, {Davies}, {Deil}, {Devin}, {Dirson},
  {Djannati-Ata{\"\i}}, {Dmytriiev}, {Donath}, {Doroshenko}, {Duffy}, {Dyks},
  {Egberts}, {Eichhorn}, {Einecke}, {Emery}, {Ernenwein}, {Feijen}, {Fegan},
  {Fiasson}, {Fichet de Clairfontaine}, {Fontaine}, {Funk}, {F{\"u}{\ss}ling},
  {Gabici}, {Gallant}, {Giavitto}, {Giunti}, {Glawion}, {Glicenstein},
  {Gottschall}, {Grondin}, {Hahn}, {Haupt}, {Hermann}, {Hinton}, {Hofmann},
  {Hoischen}, {Holch}, {Holler}, {H{\"o}rbe}, {Horns}, {Huber}, {Jamrozy},
  {Jankowsky}, {Jankowsky}, {Jardin-Blicq}, {Joshi}, {Jung-Richardt}, {Kasai},
  {Kastendieck}, {Katarzy{\'n}ski}, {Katz}, {Khangulyan}, {Kh{\`e}lifi},
  {Klepser}, {Kluzniak}, {Komin}, {Konno}, {Kosack}, {Kostunin}, {Kreter},
  {Lamanna}, {Lemi{\`e}re}, {Lemoine-Goumard}, {Lenain}, {Levy}, {Lohse},
  {Lypova}, {Mackey}, {Majumdar}, {Malyshev}, {Malyshev}, {Marandon},
  {Marchegiani}, {Marcowith}, {Mares}, {Mart{\`\i}-Devesa}, {Marx}, {Maurin},
  {Meintjes}, {Meyer}, {Moderski}, {Mohamed}, {Mohrmann}, {Montanari}, {Moore},
  {Morris}, {Moulin}, {Muller}, {Murach}, {Nakashima}, {Nayerhoda}, {de
  Naurois}, {Ndiyavala}, {Niemiec}, {Oakes}, {O'Brien}, {Odaka}, {Ohm},
  {Olivera-Nieto}, {de Ona Wilhelmi}, {Ostrowski}, {Panter}, {Panny},
  {Parsons}, {Peron}, {Peyaud}, {Piel}, {Pita}, {Poireau}, {Priyana Noel},
  {Prokhorov}, {Prokoph}, {P{\"u}hlhofer}, {Punch}, {Quirrenbach}, {Raab},
  {Rauth}, {Reichherzer}, {Reimer}, {Reimer}, {Remy}, {Renaud}, {Rieger},
  {Rinchiuso}, {Romoli}, {Rowell}, {Rudak}, {Ruiz-Velasco}, {Sahakian},
  {Sailer}, {Salzmann}, {Sanchez}, {Santangelo}, {Sasaki}, {Scalici},
  {Sch{\"u}ssler}, {Schutte}, {Schwanke}, {Schwemmer}, {Seglar-Arroyo},
  {Senniappan}, {Seyffert}, {Shafi}, {Shiningayamwe}, {Simoni}, {Sinha}, {Sol},
  {Specovius}, {Spencer}, {Spir-Jacob}, {Stawarz}, {Sun}, {Steenkamp},
  {Stegmann}, {Steinmassl}, {Steppa}, {Takahashi}, {Tavernier}, {Taylor},
  {Terrier}, {Tiziani}, {Tluczykont}, {Tomankova}, {Trichard}, {Tsirou},
  {Tuffs}, {Uchiyama}, {van der Walt}, {van Eldik}, {van Rensburg}, {van
  Soelen}, {Vasileiadis}, {Veh}, {Venter}, {Viana}, {Vincent}, {Vink},
  {V{\"o}lk}, {Wadiasingh}, {Wagner}, {Watson}, {Werner}, {White},
  {Wierzcholska}, {Wun Wong}, {Yusafzai}, {Zacharias}, {Zanin}, {Zargaryan},
  {Zdziarski}, {Zech}, {Zhu}, {Zmija}, {Zorn}, {Zouari}, \&
  {Zywucka}}]{we_ufos}
{H.~E.~S.~S. Collaboration}, {Abdallah}, H., {Aharonian}, F., {et~al.} 2021,
  arXiv e-prints, arXiv:2106.00551

\bibitem[{Hiroshima {et~al.}(2018)Hiroshima, Ando, \&
  Ishiyama}]{Hiroshima:2018kfv}
Hiroshima, N., Ando, S., \& Ishiyama, T. 2018, Phys. Rev. D, 97, 123002

\bibitem[{{H{\"u}tten} {et~al.}(2016){H{\"u}tten}, {Combet}, {Maier}, \&
  {Maurin}}]{hutten16}
{H{\"u}tten}, M., {Combet}, C., {Maier}, G., \& {Maurin}, D. 2016, \jcap, 2016,
  047

\bibitem[{{H{\"u}tten} {et~al.}(2019{\natexlab{a}}){H{\"u}tten}, {Combet}, \&
  {Maurin}}]{clumpy_v3}
{H{\"u}tten}, M., {Combet}, C., \& {Maurin}, D. 2019{\natexlab{a}}, Computer
  Physics Communications, 235, 336

\bibitem[{{H{\"u}tten} {et~al.}(2019{\natexlab{b}}){H{\"u}tten}, {Stref},
  {Combet}, {Lavalle}, \& {Maurin}}]{hutten19}
{H{\"u}tten}, M., {Stref}, M., {Combet}, C., {Lavalle}, J., \& {Maurin}, D.
  2019{\natexlab{b}}, Galaxies, 7, 60

\bibitem[{Kamionkowski {et~al.}(2010)Kamionkowski, Koushiappas, \&
  Kuhlen}]{Kamionkowski:2010mi}
Kamionkowski, M., Koushiappas, S.~M., \& Kuhlen, M. 2010, Phys.\ Rev.\ D, 81,
  043532

\bibitem[{{Li} \& {Ma}(1983)}]{1983ApJ...272..317L}
{Li}, T.~P. \& {Ma}, Y.~Q. 1983, \apj, 272, 317

\bibitem[{Mirabal {et~al.}(2016)Mirabal, Charles, Ferrara, Gonthier, Harding,
  S\'anchez-Conde, \& Thompson}]{Mirabal:2016huj}
Mirabal, N., Charles, E., Ferrara, E., {et~al.} 2016, Astrophys. J., 825, 69

\bibitem[{Mollitor {et~al.}(2015)Mollitor, Nezri, \&
  Teyssier}]{Mollitor:2014ara}
Mollitor, P., Nezri, E., \& Teyssier, R. 2015, Mon. Not. Roy. Astron. Soc.,
  447, 1353

\bibitem[{Navarro {et~al.}(1997)Navarro, Frenk, \& White}]{Navarro:1996gj}
Navarro, J.~F., Frenk, C.~S., \& White, S. D.~M. 1997, Astrophys. J., 490, 493

\bibitem[{Read(2014)}]{Read:2014qva}
Read, J. 2014, J. Phys. G, 41, 063101

\bibitem[{Sawala {et~al.}(2016)}]{Sawala:2015cdf}
Sawala, T. {et~al.} 2016, Mon. Not. Roy. Astron. Soc., 457, 1931

\bibitem[{Springel {et~al.}(2008)Springel, Wang, Vogelsberger, Ludlow, Jenkins,
  Helmi, Navarro, Frenk, \& White}]{Springel:2008cc}
Springel, V., Wang, J., Vogelsberger, M., {et~al.} 2008, Mon. Not. Roy. Astron.
  Soc., 391, 1685

\bibitem[{Steigman {et~al.}(2012)Steigman, Dasgupta, \&
  Beacom}]{Steigman:2012nb}
Steigman, G., Dasgupta, B., \& Beacom, J.~F. 2012, Phys. Rev., D86, 023506

\bibitem[{Zechlin {et~al.}(2012)Zechlin, Fernandes, Elsaesser, \&
  Horns}]{Zechlin:2011kk}
Zechlin, H.~S., Fernandes, M.~V., Elsaesser, D., \& Horns, D. 2012, Astron.
  Astrophys., 538, A93

\end{thebibliography}
